\documentclass{PoS}
\usepackage{amsmath}
\usepackage{graphicx, subfigure}
\usepackage{epsfig}
\usepackage{wrapfig}
\usepackage{amsmath}
\usepackage{amssymb}

\newcommand{\betaLC}{\beta_{\mathrm{L}}^{~\mathrm{c}}}

\newcommand{\gL}{g_{\mathrm{L}}}

\newcommand{\gLC}{g_{\mathrm{L}}^{~\mathrm{c}}}

\newcommand{\gTC}{g_{\mathrm{T}}^{~\mathrm{c}}}

\title{Chiral symmetry restoration\\ in QCD with many flavours}

\ShortTitle{Chiral symmetry restoration in QCD with many flavours}

\author{\speaker{Maria Paola Lombardo}%

\\INFN-Laboratori Nazionali di Frascati, I-00044, Frascati (RM),  and
INFN-Sezione di Pisa, I-56123, Pisa, Italy\\E-mail: \email{lombardo@lnf.infn.it}}
\author{{Kohtaroh Miura}\\
Kobayashi-Maskawa Institute for the Origin of Particles and the Universe, Nagoya University, Nagoya 464-8602,Japan\\
E-mail: \email{Kohtaroh.Miura@cpt.univ-mrs.fr}}
\author{Tiago J. Nunes da Silva, Elisabetta Pallante\\
Centre for Theoretical Physics, University of Groningen, 9747 AG, Netherlands\\E-mail: \email{tiagoj.nunes@gmail.com,e.pallante@rug.nl}}

\abstract{We discuss the phases of QCD in the parameter space spanned by the
  number of light flavours and the temperature with respect to the realisation
  of chiral and conformal symmetries.
  The intriguing interplay of these symmetries is best studied by means of 
lattice simulations, and some selected results from our recent work are presented
here.}

\FullConference{9th International Workshop on Critical Point and Onset of Deconfinement\\
   17-21 November, 2014\\
    ZiF (Center of Interdisciplinary Research), University of Bielefeld, Germany}

\begin{document}

\section{Chiral symmetry and conformal symmetry in QCD with many flavours}
In this talk, 
at a variance with the phase diagram in the temperature, chemical potential plane much discussed at this meeting, we will explore the phases of QCD varying the temperature and the number of flavours. This is of course a vast subject\cite{SCGT}, and here 
we will limit ourselves to the presentation of some of our recent results
\cite{Lombardo:2014mda,Lombardo:2014pda}.  

In general, we know that framing a theory in a larger parameter space often adds to our understanding. More specifically, if the new axis is the number of light flavours, we have the chance to
access a new regime of strong interactions:  when the number of flavours exceeds a critical number, an infra-red fixed point (IRFP) appears
and prevents the coupling from growing large
enough to break chiral symmetry. 
The theory is then  conformal invariant, and this remains true till asymptotic freedom is lost: the range of flavours between
the onset of conformality and the loss of asymptotic freedom
is known as the conformal window of strong interactions.
Inside the conformal window the running of
the coupling is regulated by two fixed point: the familiar UV fixed point at large momentum scales, and this novel IRFP for small momenta. Very interestingly,
the coupling at the IRFP can still be largish: we can learn about strongly interacting conformal theories, and we can assess up to which extent such theories
might offer a guidance to the physics of the strongly interactive quark gluon plasma.

Let us remind ourselves that (near)conformality plays quite an important role in the modelling of the plasma: bulk viscosity is set to zero by fiat in many hydrodynamics study,  and in general AdS/CFT predictions have been widely scrutinised and contrasted with experimental observations and numerical results-- a significant example
of these predictions is the low value of the viscosity to entropy ratio,
$\frac{\eta}{S} \simeq 1/4\pi$, which appears to describe well the hydrodynamic evolution of the plasma
 after the collisions.   To make more transparent a possible connection, let us then consider 
that both physics intuition and phenomenological analysis
based on functional renormalisation group~\cite{BraunGies}
and finite temperature holographic QCD \cite{EKMJ}--
\cite{Bigazzi:2014qsa}
indicate that the conformal phase of cold, many flavour QCD and 
the high temperature chirally symmetric phase are continuously connected. 
In particular, the onset of the conformal window coincides with 
the vanishing of the transition temperature, and the conformal
window appears as a zero temperature limit of
a possibly strongly interacting QGP.

From a general field theory viewpoint, the analysis of the phase diagram 
of strong interactions as a function of the number of flavours 
adds to our  knowledge  of strong interactions and
their fundamental mechanisms. From a particle phenomenology viewpoint,
this study deals with a class of models which might play a relevant role
in model building beyond the standard model,
which explain the origin of mass using strong coupling mechanisms.

\begin{figure}
\centering
\includegraphics[width=8 truecm]{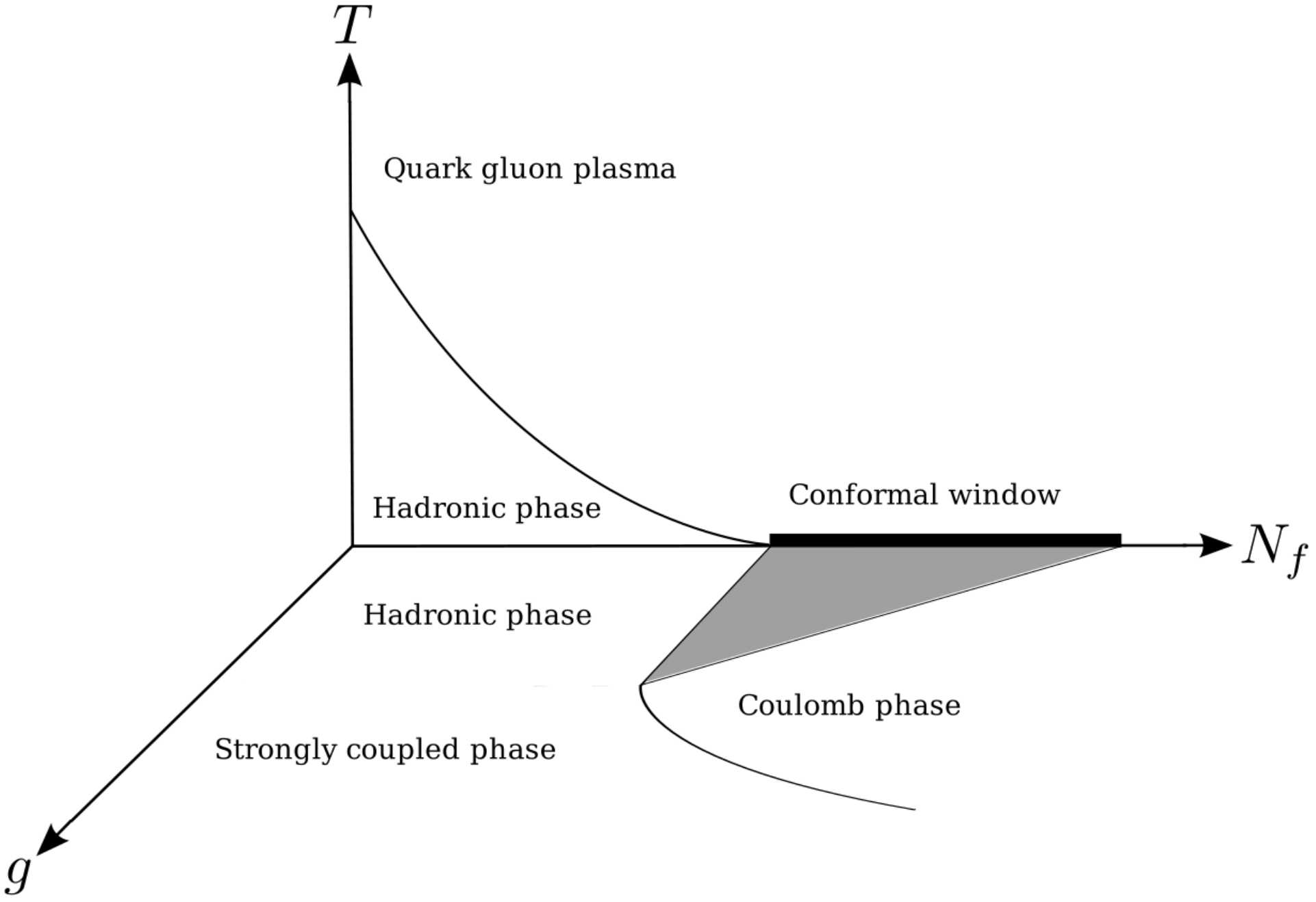}
\caption{\label{Phaseplot} A sketch of the phase diagram of
QCD-like theories in the temperature ($T$), flavour number ($N_f$) and bare
coupling ($g$) space. In the T-$N_f$ plane, the critical line is a phase
boundary between the chirally broken hadronic phase and the chirally
symmetric quark gluon plasma, the zero temperature end point of which is
the onset of the conformal window. The zero temperature projected plane is
inspired by the scenario in Refs.~\protect\cite{Miransky:1997}.}
\label{fig:Combined_phase}
\end{figure}

A simplified view of the phase diagram which we will consider here is sketched 
Figure ~\ref{fig:Combined_phase}:
the axis is simply the number of light flavours. Ordinary QCD
belongs to the hadronic phase. 
The conformal region is on the right hand side, 
and is separated by an essential singularity  from the hadronic phase.
Clearly, as in any system undergoing a phase
transition, the nature and extent of the critical window are purely
dynamical questions whose answer cannot be guessed a priori.
Since the underlying dynamics is completely non-perturbative, lattice
calculations are the only tool to perform an ab initio, rigorous
study of these phenomena\cite{Appelquist:2011dp}--\cite{deForcrand}.

We now turn to the presentation of our results. The following Section is a
concise summary of our most recent paper Ref.\cite{Lombardo:2014pda}
devoted to the analysis
of the conformal phase for a fixed value of $N_f=12$. Then,
following the presentation of our recent review 
Ref.\cite{Lombardo:2014mda}, we concentrate on the pre-conformal phase,
discussing first some of its general features, and
then focussing   on the strength of the coupling
 of the Quark Gluon Plasma as a function of $N_f$.
 We close with a brief summary.

\section{The cold conformal phase}\label{sec:spectrum}
We study the $SU(3)$ gauge theory with twelve flavours of fermions in the fundamental representation as a possible realisation of a conformal theory.
Following the phase diagram sketched in Figure 1, we consider the
Coulomb phase which appears for lattice couplings larger than the one where
the beta function vanishes. 
Our main result here is to show that the spectrum in the Coulomb phase of the system can be described in the context of a universal conformal scaling
analysis. We then provide the nonperturbative determination of the fermion mass anomalous dimension $\gamma^*=0.235(46)$ at the infrared fixed point. 
Our recent work discusses in detail how the ideal conformal
scaling law should be modified
by finite volume and other lattice artefacts. The remarkable outcome is
that all the results from different groups indicate the same anomalous dimension
once such corrections are properly taken into account.

\begin{figure}[tbp]
  \label{fig:c_noietal} 
\centering
\includegraphics[width=.60\textwidth]{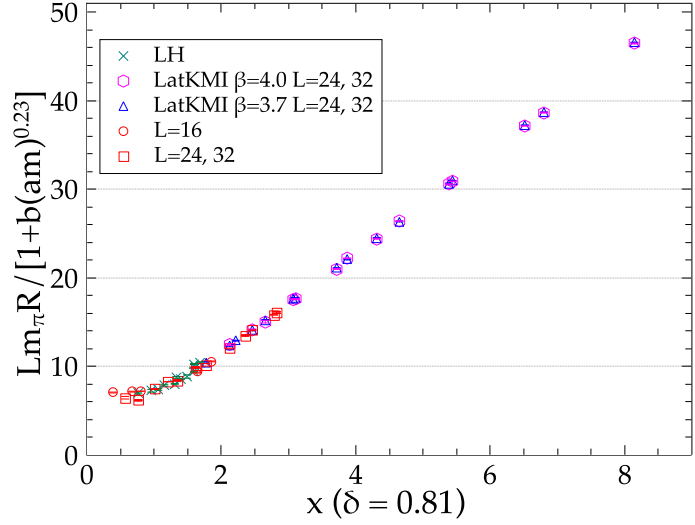}
\caption{
Collapse of curves for the rescaled pseudoscalar product $Lm_\pi R/(1+bm^\omega )$, with $\omega = 0.23$ from 4-loop perturbation theory  as a function of the universal scaling variable $x=Lm^\delta$ with $\delta=0.81$ from this work. Data are from \cite{Fodor:2011tu} at $\beta =2.2$ (LH, green crosses), \cite{Aoki:2012eq} at $\beta =4.0$ (LatKMI, magenta hexagons) and $\beta =3.7$ (LatKMI, blue triangles), and from this work for $L=24,32$ (red squares) and $L=16$ (red circles). From Ref. \cite{Lombardo:2014mda}.} 
\end{figure}

Figure~\ref{fig:c_noietal}
summarises this study, where we show the collapse on a common universal curve of data obtained with different lattice actions and lattice couplings, once perturbative corrections to the universal scaling are divided out.  Details on the
fits and optimal parameters can be found in  Ref.\cite{Lombardo:2014mda}.

This analysis leads to the determination of the mass anomalous dimension $\gamma^*$ at the IRFP. We quote the value
\begin{equation}
\label{eq:gamma}
\delta = 0.81(3)~~~~\gamma^* = 0.235(46)  
\end{equation}
This value is in agreement with the perturbative four-loop prediction, with the best-fit result of \cite{Cheng:2013xha} and not far from the first lattice determination of the fermion mass anomalous dimension for the $N_f=12$ system in \cite{Deuzeman:2009mh}, though the latter was affected by rather large uncertainties.

 To summarise, a universal power law with exponent $\delta = 0.81$ describes all would-be hadrons, with additional perturbative mass corrections of the type $1+\Delta g m^{\delta\gamma_g^*}$ in the vector and axial channels.  The pseudoscalar is the lightest state, but it is not a Goldstone boson. The vector, the scalar, the axial, and finally the nucleon follow. 
 %%%

The identification of universal contributions dictated by the conformal invariance at the fixed point and deviations from universal scaling induced in the surroundings of the fixed point has allowed for the nonperturbative determination of  the fermion mass anomalous dimension $\gamma^*=0.235(46)$ and a unified description of all lattice results for the would-be hadron spectrum of the $N_f=12$ theory.

%%%%%%%%%%%%%%%%%%%%%%%%%%%%%%%%%%%%%%%%%%%%%%%%%%%%%%%%%%%%%%%%%%%%%%
\section{Near-conformal: exploring scales}\label{sec:Tc}

In this Section we discuss results for $N_f=0,~4,~6,~8$,
approaching the conformal window from below.
In this case the results have been obtained with
a fixed bare quark mass, and no attempt has been done to
extrapolate to the chiral limit. Obviously in this case all the notions
of symmetries are approximate, due the explicit mass (chiral breaking)
term.

In order to monitor the behaviour of these theories we had
to choose some observables:  we have considered the (pseudo)critical
temperature, the 'most UV scale' available on the lattice
(which will be explained later), the zero temperature string tension, and the
$w_0$ scale\cite{Borsanyi:2012zs} defined by the Wilson
flow \cite{Luscher:2009eq} (the preliminary results for this latter observables
will not be shown here). 

\subsection{(Pseudo)critical temperature}
\begin{figure}
\centering
\includegraphics[width=10cm]{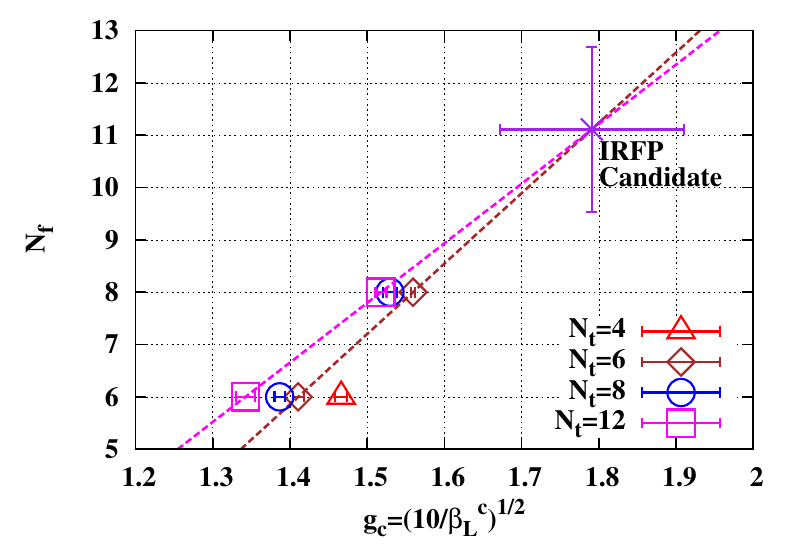}
\caption{(Pseudo) critical values of the lattice coupling
$\gLC=\sqrt{10/\betaLC}$ for theories with $N_f=6,~8$ 
and for several values of $N_t$
in the Miransky-Yamawaki phase diagram.
We have picked up $\gLC$ at $N_f = 6$ and $8$,
and considered ``constant $N_t$'' lines
with $N_t = 6,\ 12$.
If the system is still described by
one parameter beta-function in this range of coupling,
the IRFP could be located at the intersection of the
fixed $N_t$ lines -- or equivalently, in the region where
the step-scaling function vanishes. To demonstrate the procedure
--as a preliminary example -- 
we have considered the intersection
of the $N_t = 12$ and $N_t = 6$ lines}
\label{Fig:MY}
\end{figure}

We consider first 
the (pseudo)critical lattice couplings $\betaLC$ as a function
of $N_f$ and $N_t$
associated with the thermal crossover which can be observed also
with a finite mass. 
Let us plot
$\gLC(N_f,N_t) = \sqrt{10/\betaLC(N_f,N_t)}$
in the space spanned by the bare coupling $\gL$
and the number of flavour $N_f$, and 
 consider the lines which connect
$\gLC$ with $N_t$ fixed: $\gLC(N_f)|_{N_t=\mathrm{fix}}$
 (see Fig.~\ref{Fig:MY}).
 We consider the pseudo-critical lines obtained at fixed $N_t$:
the critical number of flavour $N_f^*$ can be read off
from their crossing point -- simply because this crossing point correspond
to a zero of the lattice beta function. 
In Fig.~\ref{Fig:MY} we show 
the pseudo-critical lines obtained for $N_t = 6$ and $N_t=12$
whose intersection can be estimated  at
$(\gLC, N_f^*) = (1.79\pm 0.12, 11.1\pm 1.6)$.
We emphasise that this  is a finite mass results, implying that 
$N_f^*$ is an upper limit to the chiral limit critical number of flavours. 

Let us now fix $N_f$ and consider 
the pseudo-critical temperatures $T_c$ in physical units:
\begin{align}
&T_c\equiv \frac{1}{a(\betaLC)\cdot N_t}\ .\label{eq:Tc}
\end{align}
We introduce the normalised critical temperature
$T_c/\Lambda_{\mathrm{L/E}}$ 
(see e.g. \cite{Gupta:2000hr})
where $\Lambda_{\mathrm{L}}$ ($\Lambda_{\mathrm{E}}$)
represents the lattice (E-scheme) Lambda-parameter
defined in the two-loop perturbation theory
with or without a renormalisation group inspired
improvement~\cite{CAllton}. We have observed that $T_C/\Lambda$,
regardless the prescription used,  approaches a constant by increasing $N_t$,
enabling us (with the due caveats) to interpret these asymptotic
values as continuum estimates. We can then consider 
 their $N_f$ dependence : $T_c/\Lambda$ apparently
increases with $N_f$. This suggests that $\Lambda$
vanishes faster than $T_c$ when approaching $N_f^*$, i.e.
has a strong sensitivity to the IR dynamics affected by the conformal
threshold. 

\begin{figure}
\begin{center}
\includegraphics[width=10cm]{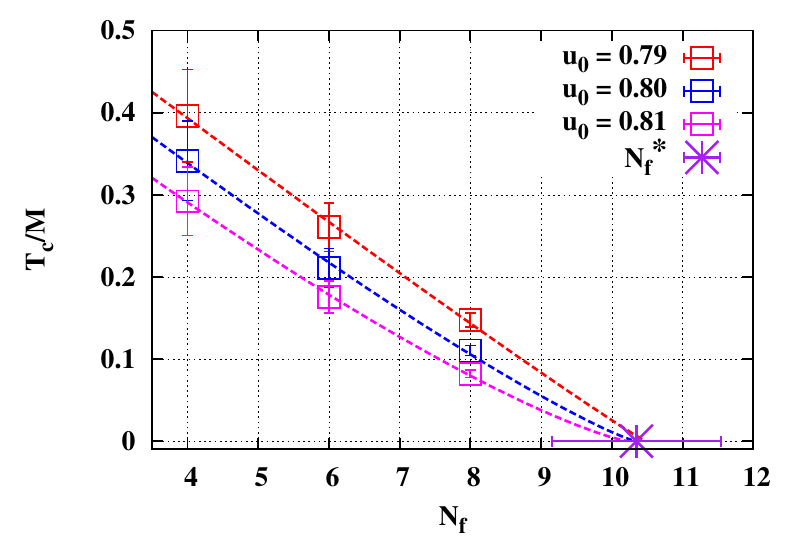}
\caption{
The $N_f$ dependence of $T_c/M$ where
$M$ is determined to be a UV scale corresponding to
different, equivalent, prescriptions and $T_c$ is the pseudocritical
temperature for a bare quark mass $m_q = 0.02$.}\label{Fig:TcM}
\end{center}
\end{figure}

We would like now to define 
a UV reference scale for different $N_f$.
We have chosen a perturbative value for the coupling
and identified -- with the help of two loop scaling -- a corresponding
mass scale which can be then used to normalise $T_c$ for different temperatures.
The outcome of this procedure is demonstrated in Figure \ref{Fig:TcM}.
On such UV scale $T_c$  is now a decreasing function of $N_f$, as expected.
Since these results have been obtained at a finite mass, the critical number of
flavours in the chiral limit is encoded in the appropriate generalisation
of the magnetic equation of state, and at least another mass value would
be needed to fix its free parameters (the normalizations).
A semiqualitative extrapolation of the data suggest that 
$T_c(m_q)$ would equal zero for
$N_f^* = 10.4 \pm 1.2$. Again this finite mass estimate provides an upper
limit to the critical number of flavours in the chiral limit.
A similar reasoning brings to the same conclusion in the case of a first
order transition. We emphasise that the logic of this exercise is not
to find a precise estimate of the critical number of flavours, rather
to provide an evidence that such a critical
number of flavours does exist,
and its approach can be observed. If and where the critical
temperature in the chiral limit reaches zero is a question
obviously beyond the scope
of an analysis performed with a fixed value of the quark mass. 

\begin{figure}
\begin{center}
\includegraphics[width=10cm]{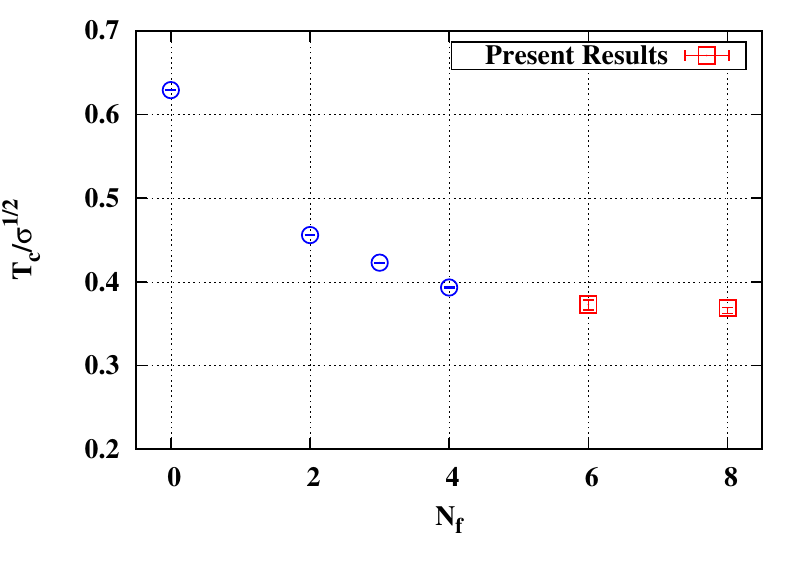}
\caption{
The $T_c/\sqrt{\sigma}$ as a function of $N_f$.
The symbol $\Box$ (red) represents the present results ($N_f = 6,8$).
For a comparison, we have quoted the $T_c/\sqrt{\sigma}$ from
\protect\cite{Laermann:1996xn} ($N_f = 0$),
\protect\cite{Karsch:2000kv} ($N_f = 2,3$),
\protect\cite{Engels:1996ag} ($N_f = 4$),
shown as $\bigcirc$ (blue) symbols. These results are still preliminary. 
}\label{Fig:Tc_s}
\end{center}
\end{figure}

Next, 
we have measured the zero temperature string tension  $\sigma$ from
Wilson Loops. 
 $\sigma$ has been  evaluated for the
same set of pseudocritical couplings we have identified in our thermal study.
%.
Because of this, we can immediately compute
$T_c / \sqrt{\sigma}$ (Figure~\ref{Fig:Tc_s}): we note 
that  the decreasing trend becomes less apparent with increasing 
$N_f$. Interestingly, this behaviour compares nicely with the findings of
a recent holographic study \cite{Bigazzi:2014qsa}.

\section{Which is the strongest coupled QGP?}\label{sec:coupling}
We consider here 
the coupling $\gTC (N_f)$ at the scale of
the (pseudo)critical temperature for each $N_f$ -- having established with
reasonable confidence asymptotic scaling we use perturbation theory
to  evolve the coupling at the scale of the lattice spacing
$a$ up to the temperature inverse scale $N_t a$.

\begin{figure}
\begin{center}
\includegraphics[width=10cm]{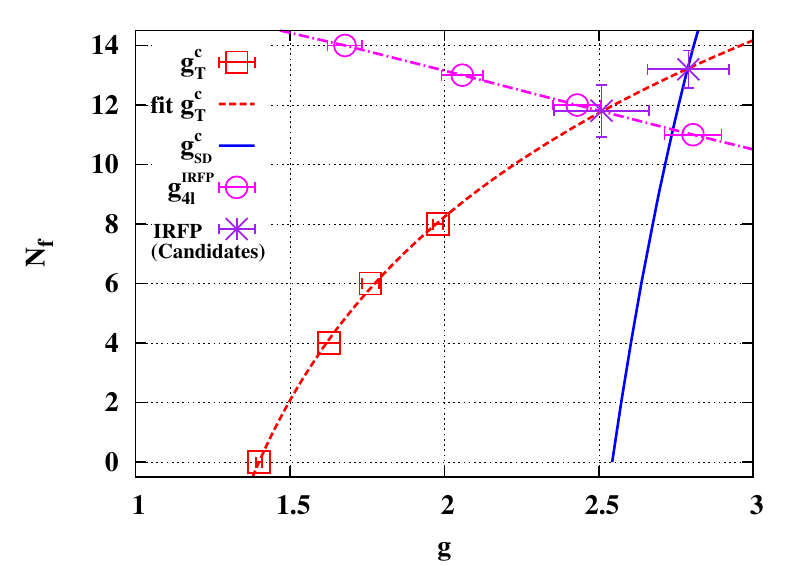}
\caption{The  couplings at $T_c$ (red $\Box$)
and the fit for them
(dashed red line, with the ansatz Eq.~(\protect\ref{eq:gTC_fit}))
and the values of the zero temperature couplings in the conformal
phase from different analytic estimates. The
thermal 
coupling increases with $N_f$ and at the critical
number of flavours it should equal the 
coupling associated with the IRFP.}
\label{Fig:gTC}
\end{center}
\end{figure}

The red ($\Box$) symbol in
Figure~\ref{Fig:gTC} shows $\gTC$ as a function of $N_f$.
We superimpose a fit obtained by using the ansatz proposed in
Ref. \cite{Liao:2012tw}

\begin{align}
{N_f(\gTC) = A\cdot \log~
\bigl[B\cdot(\gTC- \gTC|_{N_f=0}) + 1\bigr]\ .\label{eq:gTC_fit}}
\end{align}
with $A$ and $B$ fit parameters, which describes well the data. 
We note that $\gTC$ is an increasing function
of $N_f$. This indicates that the quark-gluon plasma
is more strongly coupled at larger $N_f$,
as discussed in Ref.~\cite{Liao:2012tw}.
In turn, this observation might provide a clue into the
nature of the strongly interactive quark gluon plasma.

\section{Summary}\label{sec:summary}

We have described some studies whose ultimate aim is a quantitative characterisation of the phase diagram in Figure 1. We briefly summarise here a few results: the pseudocritical temperature $T_c$ as a function of the number of light flavours, once properly normalised by use  of a UV scale, is a decreasing function
of $N_f$. The results on $T_c/\sqrt\sigma$
show a comparably milder decrease. These observations are
compatible with an (essential?)
critical behaviour and scale separation in the precritical region. The conformal
phase is strongly interactive with anomalous dimension, and the (strongly coupled) Quark Gluon Plasma is smoothly connected with such cold conformal strongly interacting phase. The latter observation might offer some motivation and guidance
for studies based on AdS/CFT correspondence and related holographic approaches.

\section{Acknowledgments}

MpL wishes to thank the organisers of CPOD2014 for a very interesting meeting
and a warm hospitality in Bielefeld. In addition, 
 the  support of the Galileo Galilei Institute
during the workshop 'Holographic Methods for Strongly Coupled Systems' is gratefully acknowledged. 
We wish to thank Marc Wagner for providing us the code for the 
Wilson loop measurements.
The numerical calculations  were performed
at CINECA in Italy,   at YITP, Kyoto University in Japan, and at KMI, Nagoya
University in Japan. 
 This work was in part based on the MILC Collaboration's 
public lattice gauge theory code~\cite{MILC}.

\end{document}